\documentclass[journal,11pt,onecolumn,draftclsnofoot]{IEEEtran}
\usepackage{ifpdf}
\usepackage[noadjust]{cite}
\usepackage[cmex10]{amsmath}
\usepackage{array}
\usepackage{dblfloatfix}
\usepackage{amssymb}
\usepackage{graphicx}
\usepackage{amsthm}
\usepackage{amsmath}
\usepackage{color}
\usepackage{makecell}

\newtheorem{theorem}{Theorem}
\def \be {\begin {eqnarray}}
\def \ee {\end {eqnarray}}

\DeclareMathOperator{\tr}{tr}
\DeclareMathOperator{\eig}{eig}
\DeclareMathOperator{\diag}{diag}
\ifCLASSINFOpdf
\else
\fi
\begin{document}
\large
\title{
Range-Spread Targets Detection in Unknown Doppler Shift via Semi-Definite Programming
}
\author{
Mai P. T. Nguyen, Iickho Song, \IEEEmembership{Fellow,~IEEE},\\
Seungwon Lee
\thanks{M. P. T. Nguyen,
I. Song, and S. Lee are
with the School of Electrical Engineering,
Korean Advanced Institute of Science and Technology, Daejeon 34141, Korea
(huongmainguyen12@gmail.com, i.song@ieee.org, kkori21@gmail.com).},
and Seokho Yoon, \IEEEmembership{Senior Member,~IEEE}
\thanks{S. Yoon is with the College of Information and Communication Engineering,
Sungkyunkwan University, 
Suwon 16419, Korea (syoon@skku.edu). 
}
}
\maketitle
\begin{abstract}
Based on the technique of generalized likelihood ratio test,
we address detection schemes for Doppler-shifted range-spread targets in Gaussian noise.
First, a detection scheme is derived 
by solving the maximization associated with the estimation of
unknown Doppler frequency with semi-definite programming. 
To lower the computational complexity of the 
detector, we then 
consider a simplification of the detector by adopting maximization
over a relaxed space.
Both of the proposed detectors  are
shown to have constant false alarm rate via numerical or theoretical analysis.
The detection performance 
of the proposed detector based on the semi-definite programming
is shown to be almost the same as that of
the conventional detector designed for known Doppler frequency.
\end{abstract}

\IEEEpeerreviewmaketitle

\newpage
\section{Introduction}
\IEEEPARstart{I}{n}
classifying problems of signal detection, we may employ 
types 
of signal and noise as a criterion. 
The types of signal include
known deterministic, unknown deterministic, and random signals, and
those of noise encompass Gaussian and non-Gaussian noise
\cite{Kay_detection}.
Regarding noise, although a non-Gaussian model better represents
experimental data and is used in some cases of practical interest \cite{Conte87},
the Gaussian model is employed more commonly due to
its validity proved by the central limit theorem and mathematical tractability.

In the meantime, while the known deterministic and random signal models arise
in a variety of communication problems, the model of deterministic signals
with some unknown parameters is commonly found in problems of
radar detection \cite{Kay_detection}. The unknown parameters in problems of
radar detection are considered as deterministic quantities or realizations of
random variables of known probability density functions (pdf).
When the unknown parameters are considered as realizations of random variables,
a detector is derived based normally on Bayesian approach. On the other hand,
the generalized likelihood ratio test (GLRT) is employed
when the unknown parameters are considered as deterministic quantities, in which
case the unknown parameters are replaced by
their maximum likelihood estimates (MLE) \cite{Kay_detection},  \cite{Bon08}.


Among radar detection problems, the detection of 
range-spread targets with an array of antennas
in noise has been addressed in a number of studies including
\cite{Gerlach99n3}--\cite{He10}, where
the target return is assumed to be known
up to a multiplication factor of the steering vector
in the cases of non-Gaussian \cite{Gerlach99n3}, \cite{Conte_Apr02}, and \cite{He10}
and Gaussian noise environment \cite{Conte01}-\cite{Gerlach00}, \cite{DeMaio_07_ortho}.
Specifically, in
\cite{Conte01} assuming the availability of the signal-free
data, called secondary data, the estimation of
the noise covariance matrix is employed in the detection, while
a modified GLRT has been proposed in \cite{Gerlach99n7}
without assuming the availability of secondary data.
In \cite{Gerlach00},
detection problem of range-spread targets has been addressed in the case of combined
thermal noise and external interference, in which the covariance matrix
of noise has the form of
an identity matrix plus an unknown positive semi-definite matrix.

The detection schemes considered in \cite{Gerlach99n3}--\cite{He10} might
suffer from a performance loss when the knowledge about the steering vector
is imperfect as in the case of mismatched steering vector.
To reduce such a detection loss when only the range of steering vector is available,
the detection problem has been addressed in several studies \cite{Fabrizio03},
\cite{Ramprashad96} under the
range models of a linear subspace 
or a cone class. 
In essence, the subspace detectors, performing the detection
by computing the energy of the measurement in the signal subspace \cite{Scharf94}
based on linear subspace models, have been considered in \cite{Scharf94}--\cite{Liu16Part1}.
In the meantime, detectors based on cone class models have been considered in
e.g., \cite{DeMaio05}--\cite{Svensson_Dec11}: In most cases of the detectors
under the cone class models,
the likelihood ratios are obtained from numerical methods, and consequently,
it is not easy 
to explain and investigate the detection nature and performance.

In this paper, we consider the problem of detecting range-spread targets
in a more general case in which the range of steering vector is unknown, a special case
of which is the case of unknown velocity of the target.
Adopting an approach similar to that in \cite{DeMaio10}, where
the detection of 
a point-like target is considered,
we first derive a GLRT based detection scheme for range-spread targets
using a semi-definite programming.
A simplified detection scheme is then obtained, which possesses
an explicit form of the likelihood ratio and thus incurs less
computational complexity. Both of the proposed detectors are shown to
not only have the constant false alarm rates (CFAR)
but also provide detection probabilities comparable to that of the conventional
detector derived with known Doppler frequency.

The rest of this paper is organized as follows.
Section II presents the problem formulation and describes
details in the derivation of the proposed detectors.
Performance analysis and comparison of the detectors
discussed in Section III. Section IV concludes this paper.

\section{Problem Formulation and Proposed Detector}
Consider the problem of detecting the presence of a range-spread target, where the
potential target is modelled by an unknown range profile (RP) and experiences a motion of unknown velocity relative
to the radar. For the detection, reflection from a sequence of $N$ identical coherent pulses
of duty cycle $T_D$ with pulse repetition interval $T_R \gg T_D$ is sampled. We assume that
the reflection of the extended target, in case a potential target appears,
is completely contained in the first $L$ range bins, whose data is referred
to as the primary data. The remaining data, called
the secondary data, occupies the remaining $K$ range bins and is composed only of
noise (by "noise" we mean radar clutter plus thermal noise). This scenario includes the special case of $L\thinspace=\thinspace1$
(i.e., target is a point scatterer) solved in \cite{Kelly86}.

Denote the sample at the $j$-th range bin for the $t$-th pulse by
$z_{tj}$ for $t = 1,2,\ldots,N$ and $j=1,2,\ldots,L+K$.
 Arranging the samples collected at range bin $j$ over $N$ consecutive pulses,
 we can form the $N \times 1$ column vector
 $\boldsymbol{z}_j \thinspace=\thinspace[z_{1j},z_{2j},\ldots,z_{Nj}]^T,$
 where $\{\cdot\}^T$ denotes the transpose of a matrix.
 The detection problem can now be stated as a problem of choosing between
 the null hypothesis
\begin{equation}\label{null_hypothesis}
H_0:  \boldsymbol{z}_j = \boldsymbol{n}_j,
j = 1,2,\ldots,L+K,
\end{equation}
and the alternative hypothesis
\begin{equation}\label{alt_hypothesis}
  H_1:
    \boldsymbol{z}_j = \left \{\begin{array}{ll}
     {\alpha}_j\boldsymbol{p}+\boldsymbol{n}_j, & j = 1,2,\ldots,L, \\
     \boldsymbol{n}_j, & j = L+1,L+2,\ldots,L+K.
    \end{array}
    \right.\
   \end{equation}
In (\ref{null_hypothesis}) and (\ref{alt_hypothesis}), the null hypothesis
$H_0$ and alternative hypothesis $H_1$ denote the cases of
noise-only and signal plus noise observations, respectively; the RP $\alpha_j$
represents the response of scatterers at range bin $j$ and is assumed constant
during the observation time; and
$\boldsymbol{p}=\left [1,\exp\left (j2\pi f_D T_R \right),\ldots ,
 \right. \linebreak \left.
\exp\left \{j2 \pi f_D(N-1)T_R\right\}\right]^{T}$
%
accounts for the Doppler shift
with $f_D$ the Doppler frequency. 
In addition,
$\boldsymbol{n}_j $ is an $N\times1$  zero-mean complex Gaussian noise vector
with the common pdf
\begin{equation}\label{normal_dist}
  f(\boldsymbol{n}_j) =
\dfrac{1}{\pi^N |\boldsymbol{C}|}\exp \left(- \boldsymbol{n}_j^H
\boldsymbol{C}^{-1}\boldsymbol{n}_j \right)
\end{equation}
 and $N \times N$ positive definite covariance matrix
$E\left[\boldsymbol{n}_j \boldsymbol{n}_j^{H}\right]=\boldsymbol{C}$
for $j = 1,2,\ldots,L+K$, where
$| \cdot |$ and $\{\cdot\}^{H}$ denote the determinant and complex conjugate
transpose of a matrix, respectively, and
$E[\cdot]$ denotes the expectation.

Assuming that the noise vectors
%
%
$\left \{ \boldsymbol{n}_i \right\}_{i=1}^{L+K}$
are independent (that is, $E\left[\boldsymbol{n}_i \boldsymbol{n}_j^{H}\right]
=\boldsymbol{0}_{N \times N}$ for $i \ne j$ with $\boldsymbol{0}_{a \times b}$
the $a \times b$ all-zero matrix)
and identically distributed
(i.i.d), the joint probability density function (pdf) of the observed data
%
%
$\left\{ \boldsymbol{z}_i \right \}_{i=1}^{L+K} $ can be expressed as
\begin{equation}\label{null_hp}
f_0\left(\boldsymbol{z}_1,\boldsymbol{z}_2,\ldots,\boldsymbol{z}_{L+K}\right)\\
\thinspace=\thinspace \dfrac{1}{\left ( \pi^N |\boldsymbol{C}|\right )^{L+K}}\exp
   \left\{-\tr \left(\boldsymbol{C}^{-1}\boldsymbol{T}_0\right)\right\}
\end{equation}
and
\begin{equation}\label{alt_hp}
  f_1\left(\boldsymbol{z}_1,\boldsymbol{z}_2,\ldots,\boldsymbol{z}_{L+K}\right) \\
\thinspace=\thinspace
  \dfrac{1}{\left ( \pi^N |\boldsymbol{C}|\right )^{L+K}}
  \exp\left\{-\tr\left(\boldsymbol{C}^{-1}\boldsymbol{T}_1\right)\right\}
   \end{equation}
under $H_0$ and $H_1$, respectively, where $ \tr(\cdot)$ 
denotes the trace
of a square matrix. In (\ref{null_hp}) and (\ref{alt_hp}),
\begin{equation}
  \boldsymbol{T}_0  = \boldsymbol{R} \left (\boldsymbol{0}_{L \times 1} \right ) +
  \boldsymbol{S}
  \end{equation}
and
 \begin{equation}
   \boldsymbol {T}_1 = \boldsymbol{R} (\pmb{\alpha})+ \boldsymbol{S},
 \end{equation}
where
$\boldsymbol{R} (\pmb{\alpha}) = \sum \limits_{j=1}^{L}(\boldsymbol{z}_j-\alpha_j
\boldsymbol{p})(\boldsymbol{z}_j-\alpha_j\boldsymbol {p})^{H}
= \left( \boldsymbol{Z}-\boldsymbol{p}\pmb{\alpha}^T \right)
\left( \boldsymbol{Z}-\boldsymbol{p}\pmb{\alpha}^T \right)^H$
represents the data in the primary range bins, the
matrix $\boldsymbol{S} = \boldsymbol{Z}_S \boldsymbol{Z}_S^H = \sum \limits_{j
= L+1}^{L+K}\boldsymbol{z}_j \boldsymbol{z}_j^{H}$ represents the secondary data,
and $\pmb{\alpha} =\left (\alpha_1,\alpha_2,\ldots,\alpha_L \right)^T$
is the $L \times 1$ vector composed of the RP
with
$ \boldsymbol{Z}\thinspace=\thinspace\left[\boldsymbol{z}_1,
\thinspace\boldsymbol{z}_2,\ldots,\thinspace\boldsymbol{z}_L \right] $ and
$\boldsymbol{Z}_S
=\left[\boldsymbol{z}_{L+1},\thinspace\boldsymbol{z}_{L+2},\ldots,
\thinspace\boldsymbol{z}_{L+K} \right]=
\left[\boldsymbol{n}_{L+1},\boldsymbol{n}_{L+2},\ldots,
 \boldsymbol{n}_{L+K}\right]$. We will eventually employ some fluctuation
models for $\pmb{\alpha}$ later in simulations to account for the variations
of the RP over the $L$ range bins.
   Clearly $\boldsymbol{R}(\pmb{\alpha})$, $\boldsymbol{S}$, $\boldsymbol{Z}$, and
   $\boldsymbol{Z}_S$ are of sizes
 $N \times N$, $N \times N$, $N \times L$, and $N \times K$, respectively.
Note that the matrix $\boldsymbol{S}$ is  positive semi-definite
Hermitian: In addition, it 
is non-singular with probability one when
$K\gg N$. Thus, the matrix $\boldsymbol{S}$ can be regarded
positive definite Hermitian without loss of generality.

From the Neyman-Pearson criterion, the likelihood ratio test (LRT)
 for the problem of choosing between $H_0$ and $H_1$ can now be expressed as
\begin{equation}\label{first_lrt}
\dfrac{\max \limits_{\boldsymbol{p}}\max \limits_{\pmb{\alpha}}\max
\limits_{\boldsymbol{C}} f_1\left(\boldsymbol{z}_1,\boldsymbol{z}_2,\ldots,
\boldsymbol{z}_{L+K}  \right)}
{\max \limits_{\boldsymbol{C}} f_0\left(\boldsymbol{z}_1,\boldsymbol{z}_2,\ldots,
\boldsymbol{z}_{L+K}  \right)}
\thinspace\mathop{\gtrless}_{H_0}^{H_1}\thinspace G,
\end{equation}
where the threshold $G$ is chosen based on the desired false alarm rate.
Due to the unavailability of $\boldsymbol{p}$, noise covariance matrix
 $\boldsymbol{C}$, RP of the target,
 we resort to a GLRT scheme to derive the LRT,
replacing nuisance parameters with their MLEs under each hypothesis. As it is well-known,
the MLE of the noise covariance matrix $\boldsymbol{C}$ under
$H_i$ is $\boldsymbol{T}_i$ for  $i=0$ and $1$ \cite{Kelly86}. Direct
substitution of the MLEs of the noise covariance matrix $\boldsymbol{C}$
into (\ref{first_lrt}) leads to
\begin{equation}\label{lrt_det}
\dfrac{|\boldsymbol{T}_0|}{\min \limits_{\boldsymbol{p}}\min \limits_{\pmb{\alpha}}
|\boldsymbol{T}_1|}  \thinspace\mathop{\gtrless}_{H_0}^{H_1}\thinspace G_1,
\end{equation} where $G_1 = \ln G$. The minimization in the
denominator over the RP vector $\pmb{\alpha}$ is then attained for
\cite{Kelly89}
\begin{equation}\label{alpha_hat}
\pmb{\hat{\alpha}}\thinspace=\thinspace
 \kappa \left ( \boldsymbol{p} \right) \left(\boldsymbol{p}^{H}\boldsymbol{S}^{-1}
 \boldsymbol{Z}\right)^{T} ,
\end{equation}
where
\begin{equation}\label{scalar_AlphaHat}
  \kappa \left ( \boldsymbol{p} \right) \thinspace=\thinspace
    \dfrac{1}{\boldsymbol{p}^{H}\boldsymbol{S}^{-1}\boldsymbol{p}}
\end{equation}
 is a positive number since $\boldsymbol{S}^{-1}$ is positive definite.
 Hence, the GLRT (\ref{lrt_det}) can be rewritten as
\begin{equation}\label{last_reviewed}
  \dfrac{|\boldsymbol{R} \left (\boldsymbol{0}_{L \times 1} \right )+\boldsymbol{S}|}
  {\min \limits_{\boldsymbol{p}} | \boldsymbol{R}(\pmb{\hat{\alpha}}) + \boldsymbol{S}|}
\thinspace\mathop{\gtrless}_{H_0}^{H_1}\thinspace G_1.
\end{equation}
After some manipulations as shown in Appendix \ref{app-b},
the GLRT is recast as
\begin{equation}\label{myGLRT}
t^{\dagger} \thinspace\mathop{\gtrless}_{H_{\mathit{0}}}^{H_{\mathit{1}}} \thinspace
G_2,
\end{equation}
where
\begin{equation}\label{t_ast}
  t^{\dagger} \thinspace = \thinspace
  \max \limits_{\boldsymbol{p}} \,
  \kappa \left ( \boldsymbol{p} \right) \boldsymbol{p}^{H}\boldsymbol{S}^{-1}\boldsymbol{Z}
\boldsymbol{X}^{-1} \boldsymbol{Z}^{H}\boldsymbol{S}^{-1}\boldsymbol{p}
\end{equation}
and $G_2$ is a suitable modification of $G_1$. In (\ref{t_ast}),
\begin{equation}\label{X_lrt}
    \boldsymbol{X}=\boldsymbol{I}_L + \boldsymbol{\Upsilon},
\end{equation}
where \be \boldsymbol{\Upsilon} =
\boldsymbol{Z}^{H}\boldsymbol{S}^{-1}\boldsymbol{Z} \ee and
$\boldsymbol{I}_L$ is the $L \times L$ identity matrix. Clearly,
$t^{\dagger} \geqslant 0$ since $
\boldsymbol{S}^{-1}\boldsymbol{Z} \boldsymbol{X}^{-1}
\boldsymbol{Z}^{H}\boldsymbol{S}^{-1} $ is positive semi-definite
and $ \kappa \left ( \boldsymbol{p} \right)$ is positive. In
passing, let us note that, when $\boldsymbol{p}$ is known, an LRT
similar to (\ref{myGLRT}) is derived in \cite{Conte01}.

\subsection{Detector based on Semi-Definite Problem}
When $\boldsymbol{p}$ is unknown, the maximization in
(\ref{t_ast}) can be solved by searching over $\boldsymbol{p}$,
which unfortunately is not quite feasible in practice. We thus
change (\ref{t_ast}) into an equivalent problem for which
efficient algorithms can be employed.

Noting that $\boldsymbol{p}=\left[1,\exp (j\theta ),\ldots
,\exp\{j(N-1)\theta\}\right]^{T}$, where $\theta = 2\pi f_D T_R
\in [0,2\pi)$ is the Doppler phase, let us first rewrite the maximization in
(\ref{t_ast}) as
\begin{equation}\label{f_theta_minimization}
\begin{array}{ll}
{\text{minimize}} & t \\
\text{s.t} & f(\theta,t)\geqslant 0,
\end{array}
\end{equation}
where
\begin{equation}\label{f_theta_def}
  f(\theta,t)=
ty_0 - x_0 + 2\Re \left\{ \sum \limits_{k=1}^{N-1}\left(ty_k - x_k\right)\exp(jk\theta)
 \right\} .
\end{equation}
In (\ref{f_theta_def}),
\begin{eqnarray}
x_k \thinspace &=& \thinspace \sum\limits_{n-m=k} \left (\boldsymbol{S}^{-1}
\boldsymbol{Z}\boldsymbol{X}^{-1}\boldsymbol{Z}^{H}\boldsymbol{S}^{-1}\right)_{mn}
\end{eqnarray}
and
\begin{equation}\label{yk_poly}
y_k \thinspace = \thinspace \sum\limits_{n-m=k} \left (\boldsymbol{S}^{-1}\right)_{mn}
\end{equation}
for $k= 0,1,\ldots,N-1$, where $(\cdot)_{mn}$ denotes the elements
at the $m$-th row and $n$-th column and $\Re \{\cdot \}$ denotes
the real part. 
It is easy to see that
 $x_0 = \tr \left(\boldsymbol{S}^{-1}\boldsymbol{Z}\boldsymbol{X}^{-1}
 \boldsymbol{Z}^{H}\boldsymbol{S}^{-1}\right)$
and $ y_0=\tr \left(\boldsymbol{S}^{-1}\right) $ are real since $
\boldsymbol{S}^{-1}\boldsymbol{Z} \boldsymbol{X}^{-1}
\boldsymbol{Z}^{H}\boldsymbol{S}^{-1} $ and $\boldsymbol{S}^{-1}$
are Hermitian and that the constraint 
%
%
$ f(\theta,t) \geqslant  0 $
is derived from, and therefore equivalent to, $ t - \kappa \left ( \boldsymbol{p} \right)
\boldsymbol{p}^{H}\boldsymbol{S}^{-1}\boldsymbol{Z}
\boldsymbol{X}^{-1}
\boldsymbol{Z}^{H}\boldsymbol{S}^{-1}\boldsymbol{p} \geqslant 0 $.

Now, observe that if $f(\theta,t) \geqslant 0 \thinspace
\text{over} \thinspace \theta \in [0,2\pi)$ then $t \geqslant t^{\dagger}$
and vice versa. In other words, the set $\left \{t:t \geqslant t^{\dagger}\right \}$
of $t$ is the same as the set $\left \{t: f(\theta,t) \geqslant 0,
\thinspace
\right. \linebreak \left.
\theta \in [0,2\pi)\right\}$ of $t$. Hence, the solution to
the problem (\ref{f_theta_minimization}) is the minimum among the
values of $t$ making $f(\theta,t)$ non-negative over $[0,2\pi)$: To find the minimum,
we apply the following theorem, modified from a theorem in \cite{Roh06}.
\begin{theorem}
The function $f(\theta,t)$ is non-negative over $[0,2\pi)$
if and only if there exists a matrix $\boldsymbol{V} \in \mathbb{H}^{N \times N}$
such that
 \begin{equation}\label{Roh_Mapply}
   t\boldsymbol{y}-\boldsymbol{x} =
 \boldsymbol{W}^H \diag \left ( \boldsymbol{W} \boldsymbol{V} \boldsymbol{W}^H \right).
 \end{equation}
Here, $\mathbb{H}^{N \times N}$ denotes the set of $N \times N$ non-negative definite
Hermitian matrices, $\boldsymbol{y} = \left [y_0, y_1,\ldots,y_{N-1}\right ]^T$,
$\boldsymbol{x} = \left [x_0, x_1,\ldots,x_{N-1}\right]^T$,
$\diag{\left ( \cdot \right )}$ denotes the $N \times 1$ vector formed
with the diagonal elements, and
 \begin{equation}\label{W_dft}
   \boldsymbol{W} \thinspace= \thinspace
   \left[\boldsymbol{w}_0,\boldsymbol{w}_1,\ldots,\boldsymbol{w}_{N-1}\right]
 \end{equation}
is an $N \times N$ matrix, where $
%
%
\boldsymbol{w}_i = \left[ 1, \omega_{M,i},\ldots, \omega_{M,i}^{N-1} \right]^T$
with $\omega_{M,i} = \exp\left (-j\dfrac{2\pi i}{M}\right)$ for a number
$M \geqslant 2N-1$ and $ i = 0,1,\ldots,N-1$.
\end{theorem}
Applying Theorem 1, the minimization (\ref{f_theta_minimization}) can finally be recast as
\begin{equation}\label{sdp_minimization}
\begin{array}{ll}
 {\text{minimize}} & t \\
                           \text{s.t} &
                           t \boldsymbol{y} - \boldsymbol{x} =
 \boldsymbol{W}^H \diag \left ( \boldsymbol{W} \boldsymbol{V} \boldsymbol{W}^H \right ) , \\
  &   \boldsymbol{V} \in \mathbb{H}^{N \times N}.
  \end{array}
\end{equation}
Since the minimization problem in (\ref{sdp_minimization}) is a
semi-definite problem (SDP), a type of convex optimization problem \cite{Roh06},
it can be solved efficiently by
various well-known methods such as 
the interior-point, first-order, and bundle methods \cite{Vandenberghe96}.

The resulting detector, which will be called the SDP detector,
eventually assumes 
the likelihood ratio test
\begin{equation}\label{t_ast_LRT}
  t_C \mathop{\gtrless}_{H_{\mathit{0}}}^{H_{\mathit{1}}} G_2,
\end{equation}
where $t_C$ 
is the solution of (\ref{sdp_minimization}) obtained by, for
example, the interior-point method.

Theoretically,
the performance of the detector (\ref{t_ast_LRT}) would be the same as that
of (\ref{myGLRT}). On the other hand,
the performance of (\ref{myGLRT}) will in practice depend on the resolution of the
searching grid of $\theta$ (or equivalently, the
unknown Doppler frequency) while the detector (\ref{t_ast_LRT})
is less dependent on the searching grid: In addition, various efficient algorithms
can be employed in solving (\ref{t_ast_LRT}) as mentioned above.
%
%
%
In passing, let us mention that the study in \cite{DeMaio10} also applied
the result in \cite{Roh06} for the detection of point-like targets
in correlated Gaussian noise under unknown direction of arrival.

\subsection{Detector based on Maximization in the Reduced Space}

The processing time to obtain the likelihood ratio $t_C$
from (\ref{sdp_minimization}) via a semi-definite program 
increases with the number of range bins (that is, with the range resolution
of the radar), as it will be shown later in
the numerical results. Another drawback of the likelihood ratio
$t_C$ is that it has no explicit form, not allowing an insight into the
performance characteristics of a detector such as the CFAR property.

We now derive a detector that can be obtained when the search
over 
$\boldsymbol{p}$ is replaced with the search over all $N \times 1$ complex vectors.
%
%
With this relaxation, we can move a few steps further in simplifying, and
explicitly showing, the structure of the detector. 
Of course, since we do not exploit \emph{a priori} knowledge about $\boldsymbol{p}$,
%
%
the simplification is achieved at the expense of some performance loss,
and therefore, the detector will have a lower detection probability
than its original version.

Firstly, as $\boldsymbol{S}^{-1}$ is Hermitian, we have
  $\boldsymbol{S}^{-1}\thinspace=\thinspace \boldsymbol{U}_{S^{-1}}^{H}
  \pmb{\Lambda}_{S^{-1}}  \boldsymbol{U}_{S^{-1}}$ from the
unitary similarity \cite{LinearAlgebra}, where $\boldsymbol{U}_{S^{-1}}$ is
  an $N \times N$ unitary matrix such that
%
%
$\boldsymbol{U}_{S^{-1}}^{H} = \boldsymbol{U}_{S^{-1}}^{-1}$
and $\pmb{\Lambda}_{S^{-1}}$ is the $N \times N$ diagonal matrix composed
of the eigenvalues $\left \{ \lambda_i\right\}^N_{i=1}$ of $\boldsymbol{S}^{-1}$.
Note that the diagonal elements $\left \{ \lambda_i\right\}^N_{i=1}$ of
$\pmb{\Lambda}_{S^{-1}}$ are all positive with probability one since
$\boldsymbol{S}^{-1}$ is
positive definite with probability one. Defining the $N \times 1$ vector
\begin{equation}\label{relaxed_steering_vec}
\boldsymbol{l} \thinspace=\thinspace
\pmb{\Lambda}_{S^{-1}}^{1/2}\boldsymbol{U}_{S^{-1}}\boldsymbol{p}
\end{equation}
and $N \times L$ matrix
\begin{equation}\label{relaxed_data}
\boldsymbol{Y} \thinspace=\thinspace
\pmb{\Lambda}_{S^{-1}}^{1/2}\boldsymbol{U}_{S^{-1}}\boldsymbol{Z},
\end{equation}
we have
%
$\boldsymbol{l}^{H}\boldsymbol{l} =
\boldsymbol{p}^H \boldsymbol{S}^{-1}\boldsymbol{p} = \dfrac{1}
{\kappa\left(\boldsymbol{p}\right) }$ and
$\boldsymbol{l}^{H}\boldsymbol{Y} =
\boldsymbol{p}^H \boldsymbol{S}^{-1}\boldsymbol{Z}$, where
$\pmb{\Lambda}_{S^{-1}}^{1/2}$ is the $N \times N$ diagonal matrix with diagonal
elements $\left\{\sqrt{\lambda_i} \right\}^N_{i=1}$. Thus, the
right-hand side in (\ref{t_ast}) can be rewritten as
\begin{equation}\label{GLRT_relaxed}
    \max \limits_{\boldsymbol{l}} \dfrac{\boldsymbol{l}^{H}\boldsymbol{Y}
    \left(\boldsymbol{I}_L+\boldsymbol{\Upsilon}\right)^{-1}
    \boldsymbol{Y}^{H}\boldsymbol{l}}{\boldsymbol{l}^{H}\boldsymbol{l}}.
 \end{equation}
Here, it should be noticed that the ratio in
(\ref{GLRT_relaxed}) without $\max \limits_{\boldsymbol{l}}$
is the Rayleigh quotient \cite{rayl-q} of the Hermitian matrix $\boldsymbol{Y}
\left(\boldsymbol{I}_L+\boldsymbol{\Upsilon}
\right)^{-1} \boldsymbol{Y}^{H}$ at $\boldsymbol{l}$: An 
important implication of this fact is that
the maximum of the ratio, or equivalently, the solution to (\ref{GLRT_relaxed})
over all $N \times 1$ vectors 
is the same as the maximum eigenvalue of
$\boldsymbol{Y}\left(\boldsymbol{I}_L+\boldsymbol{\Upsilon}
\right)^{-1} \boldsymbol{Y}^{H}$.

Now, denoting by $\eig\{\cdot \}$ the set of non-zero eigenvalues, we have
\begin{eqnarray}
  \eig \left \{\boldsymbol{Y}\left(\boldsymbol{I}_L+\boldsymbol{\Upsilon}\right)^{-1}
  \boldsymbol{Y}^{H}\right\} &=&
  \eig \left\{\left(\boldsymbol{I}_L+\boldsymbol{\Upsilon}\right)^{-1}
  \boldsymbol{Y}^{H}\boldsymbol{Y}\right\} \nonumber \\
&=& \eig \left\{\left(\boldsymbol{I}_L+\boldsymbol{\Upsilon}\right)^{-1}
                                                   \boldsymbol{\Upsilon}\right\},
\end{eqnarray}
where the first equality is based on the fact that
$\eig \left \{ \boldsymbol{A}\boldsymbol{B}\right \}=
\eig \left \{ \boldsymbol{B}\boldsymbol{A} \right \}$
 for any $m \times n$ matrix $\boldsymbol{A}$
and $n \times m$ matrix $\boldsymbol{B}$ \cite{MatrixAnalysis} and the second equality
is from $\boldsymbol{Y}^H \boldsymbol{Y}
=\left(\pmb{\Lambda}_{S^{-1}}^{1/2}\boldsymbol{U}_{S^{-1}}\boldsymbol{Z}\right)^H
\pmb{\Lambda}_{S^{-1}}^{1/2}\boldsymbol{U}_{S^{-1}}\boldsymbol{Z}
=\boldsymbol{Z}^H \boldsymbol{U}_{S^{-1}}^H \left(
\pmb{\Lambda}_{S^{-1}}^{1/2}\right)^H \pmb{\Lambda}_{S^{-1}}^{1/2}\boldsymbol{U}_{S^{-1}}
\boldsymbol{Z} =\boldsymbol{\Upsilon} $.
Since $\boldsymbol{\Upsilon} = \boldsymbol{Z}^H \boldsymbol{S}^{-1} \boldsymbol{Z}$
is Hermitian, we have the decomposition
\begin{equation}
   \boldsymbol{\Upsilon}
   \thinspace=\thinspace
   \boldsymbol{U}_{\Upsilon}^{H}\boldsymbol{\Lambda}_{\Upsilon}\boldsymbol{U}_{\Upsilon}
\end{equation}
where $\boldsymbol{\Lambda}_{\Upsilon}$ is the $L \times L$  diagonal
matrix composed of the eigenvalues
$\left \{ d_i \right \}_{i=1}^{L}$
  of $\boldsymbol{\Upsilon}$ and $\boldsymbol{U}_{\Upsilon}$ is an $L \times L$ unitary matrix.
%
%
Thus, we get
\begin{eqnarray}\label{eig_invertB}
\eig \left\{\left(\boldsymbol{I}_L+\boldsymbol{\Upsilon}\right)^{-1}
\boldsymbol{\Upsilon}\right\}
&=& \eig\left \{\left(\boldsymbol{I}_L+\boldsymbol{U}_{\Upsilon}^{H}
\boldsymbol{\Lambda}_{\Upsilon}\boldsymbol{U}_{\Upsilon}\right)^{-1}
\boldsymbol{U}_{\Upsilon}^{H}\boldsymbol{\Lambda}_{\Upsilon}
\boldsymbol{U}_{\Upsilon}\right \} \nonumber \\
&=& \eig\left\{\left[\boldsymbol{U}_{\Upsilon}^{H}\left(\boldsymbol{I}_L
+\boldsymbol{\Lambda}_{\Upsilon}\right)\boldsymbol{U}_{\Upsilon}\right]^{-1}
\boldsymbol{U}_{\Upsilon}^{H}\boldsymbol{\Lambda}_{\Upsilon}\boldsymbol{U}_{\Upsilon}
\right \}\nonumber \\
&=& \eig\left\{\boldsymbol{U}_{\Upsilon}^{H}\left(\boldsymbol{I}_L
+\boldsymbol{\Lambda}_{\Upsilon}\right)^{-1}\boldsymbol{\Lambda}_{\Upsilon}
\boldsymbol{U}_{\Upsilon}\right\} \nonumber \\
&=& \eig\left\{\left(\boldsymbol{I}_L +
\boldsymbol{\Lambda}_{\Upsilon}\right)^{-1}\boldsymbol{\Lambda}_{\Upsilon}\right\},
\end{eqnarray}
where we have used $\boldsymbol{U}_{\Upsilon}^{-1} = \boldsymbol{U}_{\Upsilon}^H$
in the third equality and the last equality is based on the fact that
 $\eig\left\{\boldsymbol{A} \right\} = \eig\left\{ \boldsymbol{B}^{-1}
 \boldsymbol{A}\boldsymbol{B} \right\} $
%
%
when $\boldsymbol{A}$ and $\boldsymbol{B}$ are square matrices of the
same size and $\boldsymbol{B}$ is invertible \cite{Lang_LinearAlgebra}.

The matrix $\left(
\boldsymbol{I}_L+\boldsymbol{\Lambda}_{\Upsilon}\right)^{-1}\boldsymbol{\Lambda}_{\Upsilon}$
is diagonal, and therefore, its eigenvalues are the same as its elements $\left
\{\dfrac{d_i}{1+ d_i}\right\}^L_{i=1}$ since $\boldsymbol{\Upsilon}$ is positive
semi-definite, where $d_i \ge 0$ for $i=1, 2, \ldots, L$.
Noting that $\dfrac{x}{1+x}$ is an increasing function of $x \ge 0$, it is easy to see
that the maximum eigenvalue of $\left(
\boldsymbol{I}_L+\boldsymbol{\Lambda}_{\Upsilon}\right)^{-1}\boldsymbol{\Lambda}_{\Upsilon}$,
that is, the maximum value of (\ref{GLRT_relaxed}) is $\dfrac{d_{\max}}{1+d_{\max}}$ with
$d_{\max} = \max \left\{ d_1,d_2,\ldots, d_L\right\}$.

The resulting detector, which we call the maximization in the relaxed space (MRS)
detector, is thus based on the likelihood ratio test
\begin{equation}\label{ray_based_LRT}
  \dfrac{d_{\max}}{1+d_{\max}}
  \thinspace\mathop{\gtrless}_{H_{\mathit{0}}}^{H_{\mathit{1}}} \thinspace \tilde{G}_2,
\end{equation}
where $\tilde{G}_2$ is the 
threshold. In Appendix \ref{app-c}, 
it is shown that the MRS detector (\ref{ray_based_LRT}) possesses the CFAR property.

\section{Numerical Results}
In this section, we assess and compare the performance, the probabilities of
detection and false alarm, of the proposed detectors (\ref{t_ast_LRT}) and (\ref{ray_based_LRT})
with that of
the conventional detector
 \begin{equation}\label{1S_GLRT}
  \frac{\boldsymbol{p}^H\boldsymbol{S}^{-1}\boldsymbol{Z}
\boldsymbol{X}^{-1} \boldsymbol{Z}^{H}\boldsymbol{S}^{-1}\boldsymbol{p}}
        {\boldsymbol{p}^H\boldsymbol{S}^{-1}\boldsymbol{p}}
        \thinspace\mathop{\gtrless}_{H_{\mathit{0}}}^{H_{\mathit{1}}} \thinspace \tilde{G}_3
\end{equation}
derived when $\boldsymbol{p}$ is known, where $\tilde{G}_3$ is a threshold.
The detector represented by (\ref{1S_GLRT}) will be called the
one-step GLRT (OS-GLRT) \cite{Conte01}.
In short, (\ref{1S_GLRT}) is an ideal detector, of which the performance represents
the bound of other detectors designed for unknown $\boldsymbol{p}$.
\subsection{Parameters}
The detection performance of three detectors was assessed with $N = 8$ identical coherent pulses transmitted with
the pulse repetition interval $T_R = 40$ $\mu$s and Doppler frequency
$f_D = 10$ kHz (which corresponds to a velocity of $150$ m/s
for a wavelength of $\lambda = 3$ cm).
The number $L$ of range cells in the primary data is lower bounded
by the ratio of the range extent
of a target to the range resolution of the radar and it could be of several hundreds
\cite{Conte01}. To alleviate the computational burden,
we chose $L=8,12,16$, and $20$ with the energy distribution among scatterers
as shown in Table I. The size of the secondary data is chosen
as $K = 16N=128$ so that the matrix $\boldsymbol{S}$ of the secondary data is non-singular.
For use in (\ref{W_dft}), we have chosen $M = 2N -1 = 15$.
%
%
Finally, we use the software CVX (http://cvxr.com/)
to solve the semi-definite problem (\ref{sdp_minimization}) on a computer
equipped with a 3.4 GHz Intel processor.

\begin{table*}
\renewcommand{\arraystretch}{1.0}
\caption{Energy distribution at discrete scatterer locations when $L=8$, $12$, $16$, and $20$} \label{time_comp}
  \centering
      \begin{tabular}{|l|llllllllllllllllllll|}
     \hline
 $L = 8 $&  $0$ & $\dfrac{1}{16}$ & $0$ & $\dfrac{1}{2}$ & $\dfrac{1}{4}$ & $\dfrac{1}{16}$ & $\dfrac{1}{8}$ & 0
     &  &  &  &  &  &  &  & &  & &  & \\
     \hline
 $L= 12$ & $0$ & $\dfrac{1}{16}$ & $\dfrac{1}{16}$ & $0$ & $0$ & $0$ & $\dfrac{1}{16}$ & $\dfrac{1}{16}$
     & $\dfrac{1}{2}$ & $\dfrac{1}{4}$ & $0$ &$0$ &  &  &  & &  & &  &  \\
    \hline
  $L = 16$ & $0$ & $0$ & $\dfrac{1}{32}$ &$\dfrac{1}{32}$ & $0$ &$\dfrac{1}{2}$& $\dfrac{1}{4}$ & $0$
     & $\dfrac{1}{32}$& $0$& $\dfrac{1}{16}$ & $\dfrac{1}{32}$ & $0$ & $\dfrac{1}{16}$ & $0$ & $0$&  & &  &  \\
    \hline
  $L = 20$ & $0$ & $\dfrac{1}{10}$ & $0$ &$ \dfrac{1}{10}$ & $0 $& $\dfrac{1}{10}$ & $0$ &$ \dfrac{1}{10}$
    & $0$ & $\dfrac{1}{10}$ & $0$ &$\dfrac{1}{10}$ & $0$ & $\dfrac{1}{10}$ & $0$ & $\dfrac{1}{10}$ & $0$ & $\dfrac{1}{10}$ & $0$ & $\dfrac{1}{10}$\\
    \hline
      \end{tabular}
  \end{table*}

\subsection{Simulation Results and Discussion}
The performance of the SDP and MRS detectors are assessed
for the detection of
steady and fluctuating targets
embedded in Gaussian noise of various correlation degrees
 in comparison with
the OS-GLRT detector.
In addition,
the complexities of the three detectors are considered in terms of the
average processing time.
Since it is not plausible to derive explicit forms of the
false alarm and detection probabilities, we resort to Monte Carlo trials, in which
we set the false alarm rate at $10^{-4}$.
Since
the MRS and OS-GLRT detectors possess CFAR property as shown in Appendix B
and in \cite{Conte01}, respectively,
the thresholds of the MRS and OS-GLRT detectors are determined once.
The threshold of the SDP detector are, on the other hand, determined for each noise covariance.
\subsubsection{Threshold of the SDP detector}
We
first investigate if the SDP detector
also has a CFAR property.
Here, correlated noise vectors with
covariance matrix
\begin{equation}\label{cov_matrix}
 \boldsymbol{C} \thinspace = \thinspace \sigma_{n}^{2}
\begin{bmatrix}
1 & \rho &  \ldots & \rho^{N-1} \\
\rho & 1 & \ldots  & \rho^{N-2}\\
\vdots & \vdots & \ddots & \vdots \\
\rho^{N-1} & \rho^{N-2} &  \ldots & 1
\end{bmatrix}
\end{equation}
are employed, where
$\sigma_{n}^{2}$ and $\rho$ denote the average noise power at one range cell
and the one-lag correlation coefficient, respectively.

\begin{figure}
   \includegraphics[width=\linewidth]{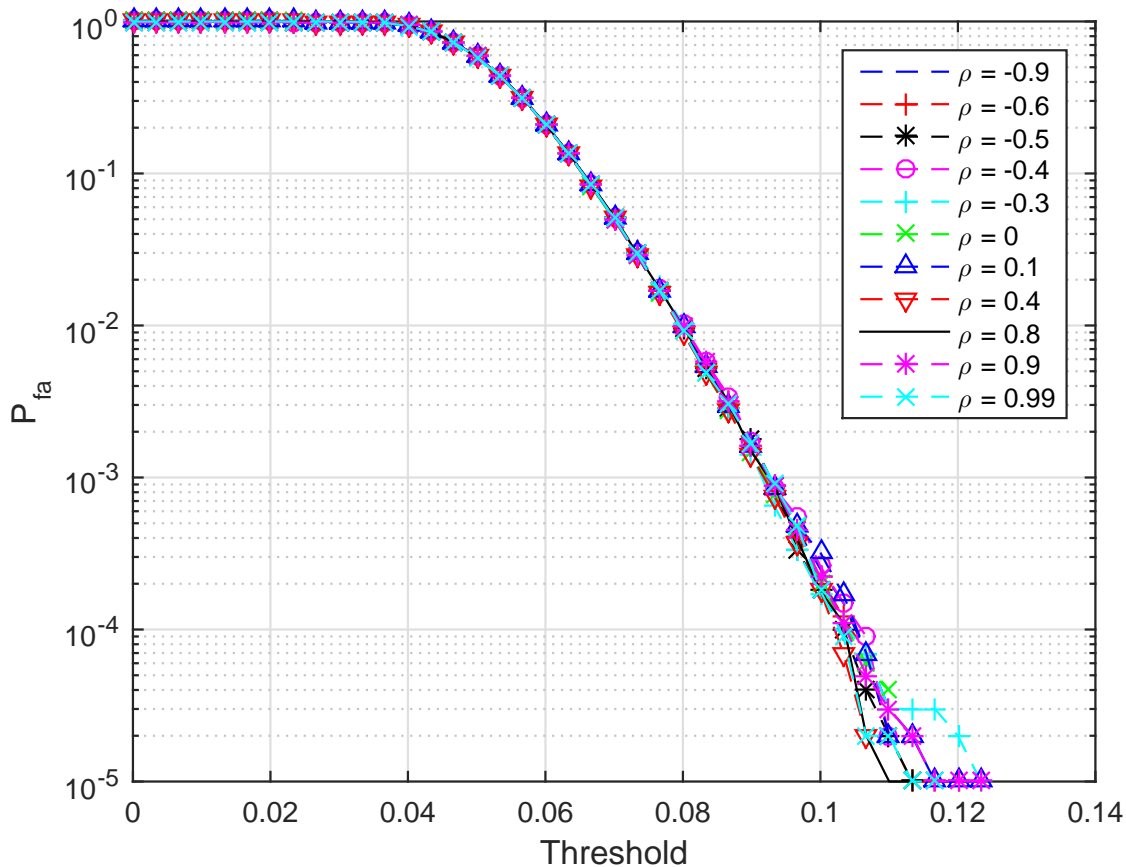}
   \caption{False alarm probability versus thresholds of the SDP detector.}
   \label{pfa_cvx_corr}
 \end{figure}

Fig. \ref{pfa_cvx_corr} shows the
false alarm rates $P_{fa}$ of the SDP detector
as  a function of threshold
at various values of $\rho$
when $\sigma_{n}^{2}= 1$. 
It is observed from the figure that
the threshold
for a specific value of $P_{fa}$
does not depend on 
 $\rho$.
We believe this observation
is a natural consequence since
the likelihood ratio of the SDP detector is a close approximation to that of
the OS-GLRT detector, which possesses a CFAR property.
\subsubsection{Detection performance with steady targets}

\begin{figure}
   \includegraphics[width=\linewidth]{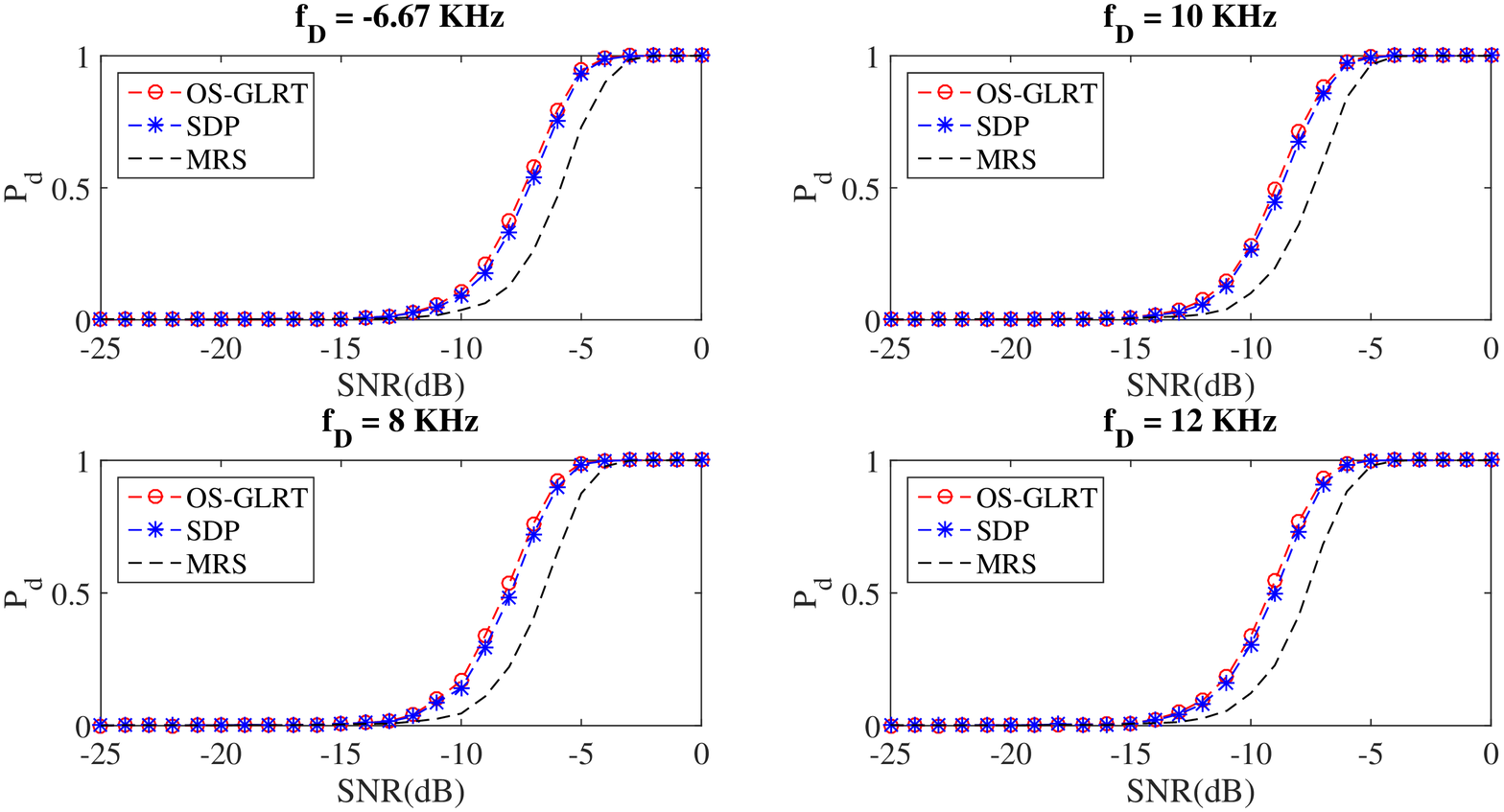}
   \caption{Detection probability of the SDP and MRS versus OS-GLRT detectors,
   steady target.}
   \label{3det-varied-fD}
 \end{figure}

We now evaluate the detection performance of proposed detectors
when the RP of a potential target
remains constant
during the observation time. 
The performances are investigated at velocities
$v = -100$, $120$, $150$, and $180$ m/s
corresponding to Doppler frequencies
$f_D = -6.67$, $8$, $10$, and $12$ KHz, respectively.
In the evaluation, we assume $L = 8$ and $\rho = 0.4$.
The detection probabilities $P_d$ are shown in Fig. \ref{3det-varied-fD}
as a function of the
signal-to-noise ratio (SNR) defined as
\begin{equation}\label{snr-std}
  \mbox{SNR} = \frac{ \| \boldsymbol{p} \|^2\sum \limits_{i = 1}^{L} | \alpha_i |^2}
               {LN \sigma_{n}^{2}}.
\end{equation}
It is observed from Fig. \ref{3det-varied-fD} that the
SDP detector achieves nearly the same performance as the OS-GLRT detector
  and outperforms the MRS detector.
For example, the SDP detector provides a gain of around $1.4$ dB over
the MRS detector for all the values of $f_D$.
To explain, we recall that information about
 the form of a target's return
$\boldsymbol{p}=\left [1,\exp\left (j2\pi f_D T_R \right),\ldots ,
\exp\left \{j2 \pi f_D(N-1)T_R\right\}\right]^{T}$
is exploited in the SDP detector; yet
to the MRS detector
such information is not exploited.

\begin{figure}
   \includegraphics[width=\linewidth]{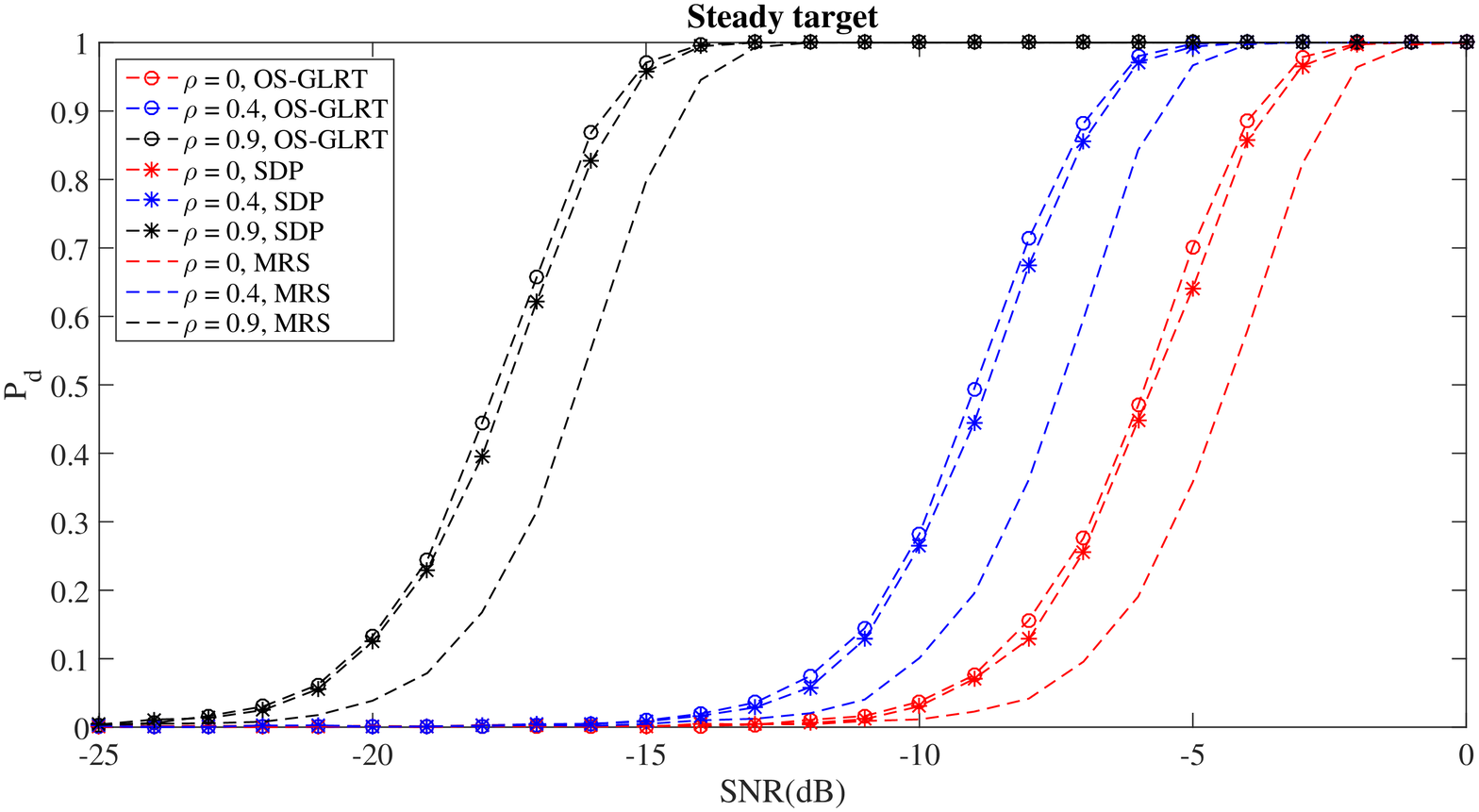}
   \caption{Detection probability of the SDP and MRS versus OS-GLRT detectors,
   steady target.}
   \label{std-3det}
 \end{figure}

Assume $L = 8$, we
 next assess detection probabilties in noise with various $\rho$:
 $0$, $0.4$, and $0.9$.
It is shown from
Fig. \ref{std-3det} that with proposed detectors,
a higher value of
$\rho$ associates to a better detection performance.
 For example, at $\mbox{SNR} = -10$ dB, with $\rho = 0.9$
 proposed detectors yield $P_d = 1$; yet with $\rho = 0.4$
 the SDP detector yields $P_d \approx 0.3$
and the MRS detector yields $P_d \approx 0.1$. 
When $\rho$ approaches $1$,
for example, $\rho = 0.999$
 (for brevity simulation result are not presented),
 all detectors achieve $P_d = 1$ at $\mbox{SNR} = -25$ dB.
 This observation is intuitively reasonable
  since a higher
   value of $\rho$ implies more similarity
between noise samples, i.e., more imformation about noise,
which possibly leads to a better detection performance.

\subsubsection{Detection performance with fluctuating targets}
Let us next address
detection performances of
proposed detectors when a potential target's RP fluctuates.
Firstly, we describe how we generate a fluctuating RP.
We note that
target's radar cross section (RCS)
representing the
total energy reflected 
from the target
also fluctuates with a fluctuating RP target.
In addition, such fluctuating RCS is well described by the
Swerling models \cite{radar-handbook}.
Based on \cite{Hughes83}, we derive
the relationship between a target's RCS, denoted by
$\sigma$, and its RP
\begin{equation}\label{RCS_RP}
 \sigma \thinspace \propto \thinspace \frac{1}{2} \sum \limits_{j=1}^{L}|\alpha_j|^2.
\end{equation}
We now assume that, for simplicity,
all 
main scatterers on the target
follow a same fluctuating pattern.
With this assumption and existing models of fluctuating $\sigma$,
the generation of a fluctuating RP is straightforward.
We employ, in the evaluation,
Swerling II and IV models, and we use $L = 8$.
Note that
Swerling II and IV models are
chi squared distributions of $2$ and $4$ degrees of freedom, respectively.
Therefore,
a Swerling II variable
exhibits larger covariance, i.e. more fluctuation, than a Swerling IV variable
at a same expectation,
In the sequel, we
refer a target with Swerling II or IV RCS
as a Swerling II or IV target, respectively. 
The averaged
SNR is defined as
 \begin{equation}
\mbox{SNR}_{av}  = \frac{E \left [\| \boldsymbol{p} \|^2\sum \limits_{i = 1}^{L} | \alpha_i |^2 \right ]}
               {LN \sigma_{n}^{2}}
\propto \frac{2\bar{\sigma} \| \boldsymbol{p}\|^2}
{LN \sigma_{n}^{2}},
 \end{equation}
 where
$av$ stands for `averaged' and $\bar{\sigma}$ the average of a target's RCS.

\begin{figure}
   \includegraphics[width=\linewidth]{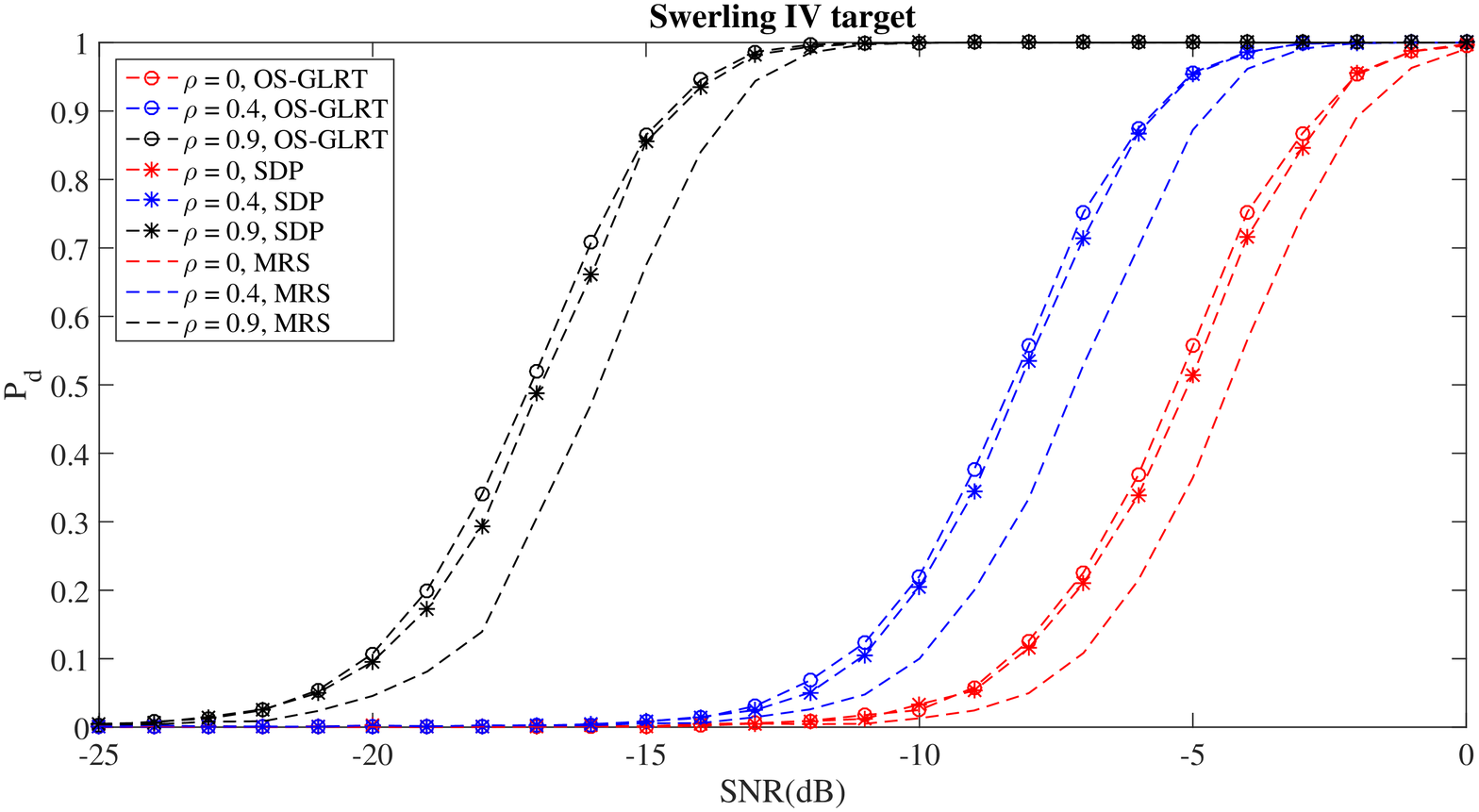}
   \caption{Detection probability of the SDP and MRS versus OS-GLRT detectors,
   Swerling IV target.}
   \label{swer4-3det}
 \end{figure}

\begin{figure}
   \includegraphics[width=\linewidth]{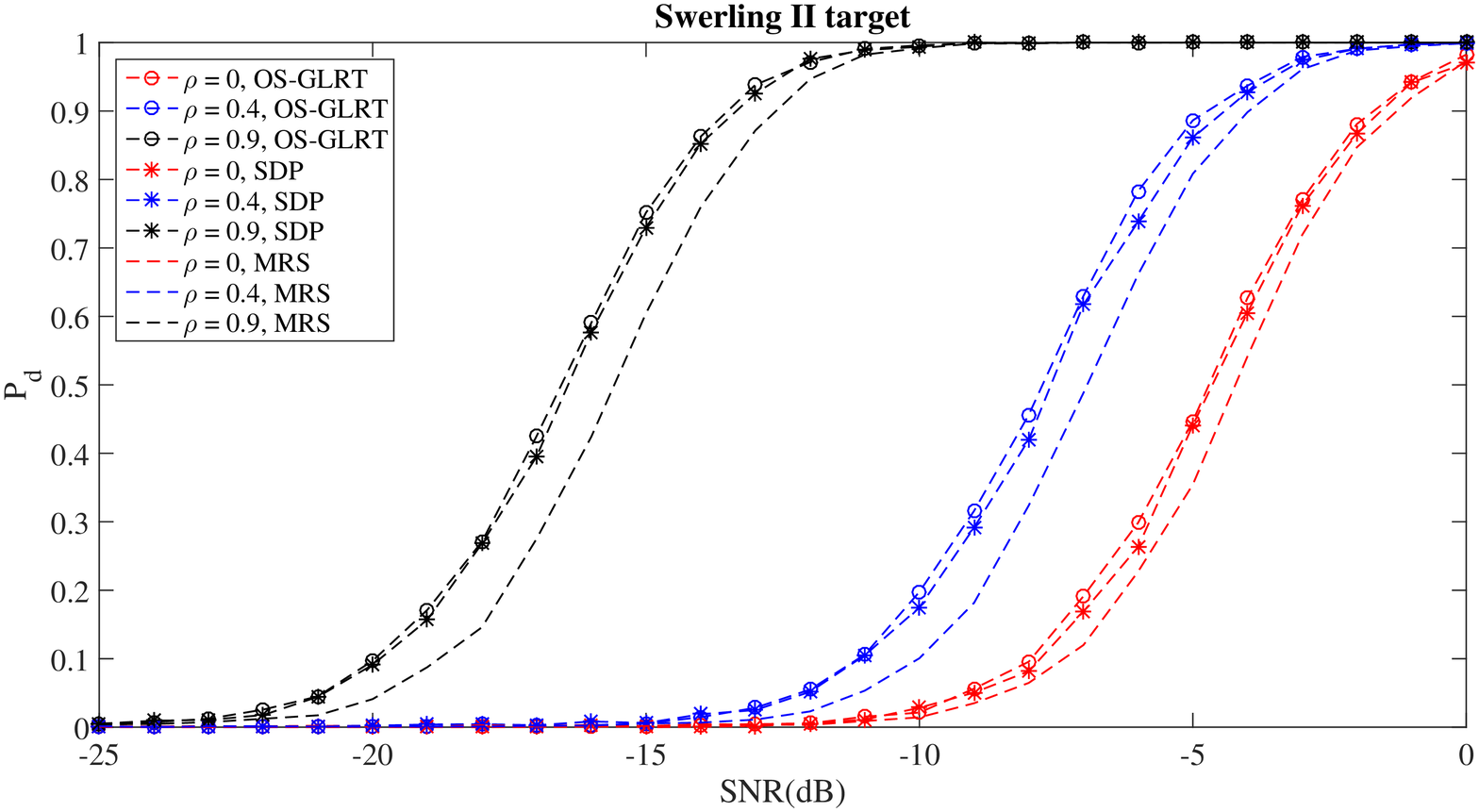}
   \caption{Detection probability of the SDP and MRS versus OS-GLRT detectors,
   Swerling II target.}
   \label{swer2-3det}
 \end{figure}

In Figs. \ref{swer4-3det} and \ref{swer2-3det}, we show
detection probabilities $P_d$
of proposed detectors, in comparison with the OS-GLRT detector,
as functions of the averaged
SNR.
These figures highlight observations identical to those derived from
Figs \ref{3det-varied-fD} and \ref{std-3det}.

The order of performance in terms of target's type
are steady, Swerling IV, and Swerling II.
Lower probabilities in the detection of a fluctuating target,
compared to a steady target, is expected
since fluctuation in a target's RP is not considered in the design stage of the three detectors.
Similarly, since a Swerling II target's return exhibits more fluctuation than that of a Swerling IV target,
detection probability with
a Swerling II target is lower than that
with a Swerling IV target.

\subsubsection{Detection with incorrect information about Doppler frequency}

\begin{figure}
   \includegraphics[width=\linewidth]{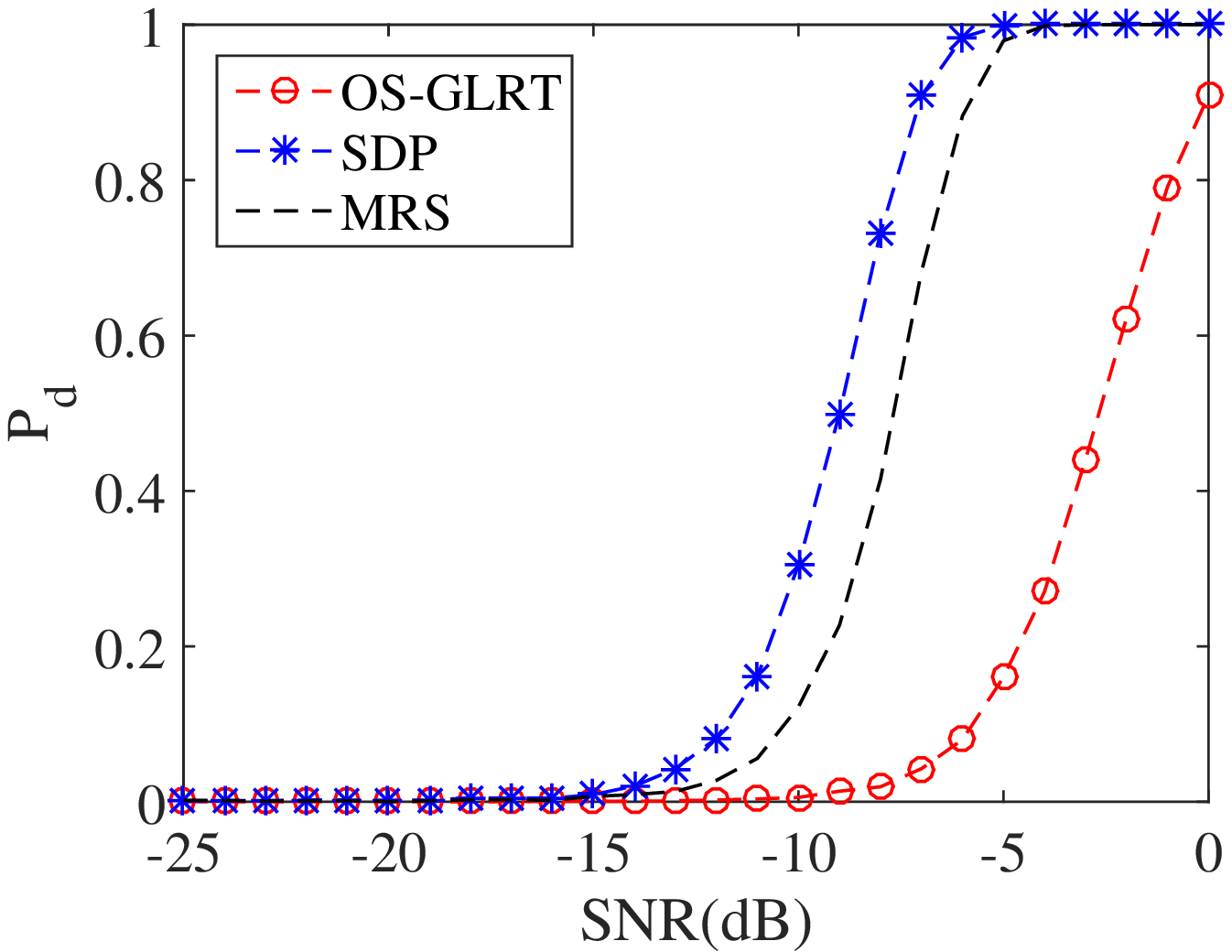}
   \caption{Detection probability of the SDP and MRS versus OS-GLRT detectors,
   Steady target when information about target's velocity is not correct.}
   \label{os-glrt-mismatch}
 \end{figure}

In Fig. \ref{os-glrt-mismatch}, we compare the
detection performance of the proposed detectors with that of the OS-GLRT detector
when information about Doppler frequency is incorrect.
In the evaluation,
the pre-assumed Doppler frequency is $10$ KHz, yet
the actual value is $12$ KHz.
 We use $L = 8$, $\rho = 0.4$, and a steady RP.
We observe that the SDP and the MRS detectors outperform
 the OS-GLRT detector
 with performance gaps, for example at SNR = $-10$ dB, of about $6$ and $5$ dB, respectively.
 This observation is easily anticipated since the operations of the SDP and the MRS detectors do not
 require Doppler frequency information; however,
 the OS-GLRT detector requires an exact information about Doppler frequency.
 \subsubsection{Complexity of proposed detectors}
\begin{table*}
\renewcommand{\arraystretch}{1.0}
\caption{Average Computational time (in second) of OS-GLRT, MRS, and SDP
detectors}\label{time_cmp_MRSvsSDP}
  \centering
  \begin{tabular}{|lllll|}
         \hline
    & $L = 8$ & $L = 12$ & $L = 16$ & $L = 20$ \\
     \hline
    OS-GLRT  & $2.44 \times 10^{-4}$ & $2.85 \times 10^{-4}$ & $3.32 \times 10^{-4}$ & $3.88 \times 10^{-4}$ \\
     \hline
    MRS  & $3.55 \times 10^{-4}$ & $3.63\times 10^{-4}$ & $4.87 \times 10^{-4}$ & $6.12 \times 10^{-4}$ \\
    \hline
    SDP  & $1.53 \times 10^{-1}$ & $1.59 \times 10^{-1}$ & $2.01 \times 10^{-1}$ & $2.03 \times 10^{-1}$ \\
        \hline
      \end{tabular}
  \end{table*}

Finally, we consider the complexity of proposed detectors in terms of
average time consumption (in second).
Here, we assess time consumption with different RP's length
$L = 8$, $12$, $16$, and $20$. For each detector,
computational time is averaged over $1000$ repetitions
to make results reliable.

It is shown in Table \ref{time_cmp_MRSvsSDP},
as easily anticipated,
 that the SDP detector assumes much more processing time,
which is in an order of $10^{-1}$, than the MRS and the OS-GLRT detectors, whose processing time are
in the order of $10^{-4}$.
Besides the computation shared by all detectors,
the MRS detector computes the
 maximum eigenvalue of $\boldsymbol{Z}^H \boldsymbol{S}^{-1} \boldsymbol{Z}$,
  which exhibits a little more computation than
 the matrix multiplications $\boldsymbol{p}^H\boldsymbol{S}^{-1}\boldsymbol{Z}
\boldsymbol{X}^{-1} \boldsymbol{Z}^{H}\boldsymbol{S}^{-1}\boldsymbol{p}$
and $\boldsymbol{p}^H\boldsymbol{S}^{-1}\boldsymbol{p}$
implemented in the OS-GLRT detector.
For the SDP detector, to implement the semi-definite programming (\ref{sdp_minimization})
exhibits even much more computation.

We notice that the SDP detector's computational complexity
is unsuitable for a radar scan mode.
Yet, though possessing fast implementation, the MRS detector may miss
a target due to lower detection performance, for which a measure is to
increase transmitted power.




\section{Conclusion}
We have proposed GLRT-based detectors
for a range-spread target of unknown Doppler frequency
embedded in Gaussian noise.
Regarding the SDP detector, we resort to a semi-definite programming to solve
the optimization problem associated to the MLE of
 unknown Doppler frequency.
To reduce the implementation complexity of the SDP detector,
we then address a suboptimal detector
by solving the above optimization over a relaxed space.
Performance of proposed detectors have been assessed,
in comparison with the conventional detector derived with known Doppler frequency,
in noise of various correlation degrees and with steady and fluctuating targets.
It is observed that the detection gap between the SDP detector
and the conventional detector is negligible.
In addition, performance of the MRS detector could be considered
as a lower bound for GLRT-based detectors when no information about a potential target
is available.
Both proposed detectors have been proved, theoretically or numerically, to possess CFAR.

The  main drawback of this paper is the lack of closed forms for
detection and false alarm probabilities of the proposed detectors, which
is for a future research.
Also, we plan to consider the detection problem under a
more generalized noise model, the spherically invariant random process.

\appendices
\section{Proof of the proposed GLRT (\ref{myGLRT}) }\label{app-b}

Let us first show the equality
\begin{equation}\label{ZandAlphaHat}
  \left(\boldsymbol{Z}-\boldsymbol{p}\pmb{\hat{\alpha}}^T\right)^{H}\boldsymbol{S}^{-1}
   \left(\boldsymbol{Z}-\boldsymbol{p}\pmb{\hat{\alpha}}^T\right)
    \thinspace=\thinspace
    \boldsymbol{\Upsilon} -  \kappa \left ( \boldsymbol{p} \right)
    \boldsymbol{u}\boldsymbol{u}^H,
    \end{equation}
where
\begin{equation}\label{def-bu}
\boldsymbol{u}=\boldsymbol{Z}^{H}\boldsymbol{S}^{-1}\boldsymbol{p}.
\end{equation}
Firstly, if we expand the left-hand side of (\ref{ZandAlphaHat}),
we have
\be \label{Zalpha_hat} \left(
\boldsymbol{Z}-\boldsymbol{p}\pmb{\hat{\alpha}}^T \right )^H
\boldsymbol{S}^{-1} \left(
\boldsymbol{Z}-\boldsymbol{p}\pmb{\hat{\alpha}}^T \right )
 &=&
\boldsymbol{\Upsilon} - \pmb{\hat{\alpha}}^{\ast}
\boldsymbol{p}^{H}\boldsymbol{S}^{-1}\boldsymbol{Z}
-\boldsymbol{Z}^{H}\boldsymbol{S}^{-1}\boldsymbol{p}\pmb{\hat{\alpha}}^T
\nonumber \\
&&+ \dfrac{1}{\kappa \left ( \boldsymbol{p} \right)}
\pmb{\hat{\alpha}}^{\ast} \pmb{\hat{\alpha}}^T \ee
using (\ref{alpha_hat}) and (\ref{scalar_AlphaHat}). Next,
since $\kappa \left ( \boldsymbol{p} \right)$ 
is a real number and $\boldsymbol{S}^{-1}$ is Hermitian, we have
$\pmb{\hat{\alpha}}^{\ast} =
 \kappa \left ( \boldsymbol{p} \right) \boldsymbol{Z}^{H}\boldsymbol{S}^{-1}\boldsymbol{p}
 = \kappa \left ( \boldsymbol{p} \right) \boldsymbol{u}
$ from (\ref{alpha_hat}) and (\ref{def-bu}). Thus, the second,
third, and last terms in the right-hand side of (\ref{Zalpha_hat})
can be written as
\begin{equation}\label{secondterm}
\pmb{\hat{\alpha}}^{\ast}\boldsymbol{p}^{H}\boldsymbol{S}^{-1}\boldsymbol{Z}
\thinspace=\thinspace \kappa \left ( \boldsymbol{p} \right)
\boldsymbol{u}\boldsymbol{u}^H,
\end{equation}
\begin{equation}\label{thirdterm}
  \boldsymbol{Z}^{H}\boldsymbol{S}^{-1}\boldsymbol{p}\pmb{\hat{\alpha}}^T
  \thinspace = \thinspace
  \boldsymbol{u}\pmb{\hat{\alpha}}^T,
\end{equation}
and
\begin{equation}\label{lastterm}
  \dfrac{1}{\kappa \left ( \boldsymbol{p} \right)} \pmb{\hat{\alpha}}^{\ast} \pmb{\hat{\alpha}}^T
    \thickspace = \thinspace
   \boldsymbol{u}\pmb{\hat{\alpha}}^T ,
\end{equation}
respectively. Using (\ref{secondterm})-(\ref{lastterm}) in
(\ref{Zalpha_hat}), we can obtain (\ref{ZandAlphaHat}).

Now, since
$\left | {\boldsymbol{A}\boldsymbol{B}}\right | \thinspace=\thinspace
\left | {\boldsymbol{A}}\right | \left | {\boldsymbol{B}}\right | $
for square matrices $\boldsymbol{A}$ and $\boldsymbol{B}$ of the same size
\cite{LinearAlgebra} and
$\left | \boldsymbol{I}_m + \boldsymbol{A}\boldsymbol{B}\right | \thinspace=\thinspace
\left | \boldsymbol{I}_n + \boldsymbol{B}\boldsymbol{A}\right | $
for any $m \times n$ matrix $\boldsymbol{A}$ and $n \times m$
matrix $\boldsymbol{B}$ \cite{MatrixAnalysis},
the numerator in the left-hand side of (\ref{last_reviewed})
can be written as \be\label{det_R0}
  \left | \boldsymbol{R}\left (\boldsymbol{0}_{L \times 1} \right ) + \boldsymbol{S} \right |
   &=& \left | \boldsymbol{S}\left(\boldsymbol{I}_{N}+
                             \boldsymbol{S}^{-1}\boldsymbol{Z}\boldsymbol{Z}^{H}\right)\right |
                             \nonumber  \\
   &=& \thinspace \left | \boldsymbol{S} \right |
   \left | \boldsymbol{I}_L+\boldsymbol{\Upsilon}\right |
   \nonumber \\
   &=& \left| \boldsymbol{S}\right | \left | \boldsymbol{X} \right | .
%
\ee
Following similar
steps, the term $\left |
\boldsymbol{R}(\pmb{\hat{\alpha}})+\boldsymbol{S}\right |$ in the
denominator in the left-hand side of (\ref{last_reviewed}) can be
expressed as  \be \label{det_R1plusS} \left |
\boldsymbol{R}(\pmb{\hat{\alpha}})+\boldsymbol{S}\right | &=&
\left |
\boldsymbol{S}+\left(\boldsymbol{Z}-\boldsymbol{p}\pmb{\hat{\alpha}}^T\right
)
  \left(\boldsymbol{Z}-\boldsymbol{p}\pmb{\hat{\alpha}}^T\right)^{H}\right|  \nonumber \\
   &=& \left |  \boldsymbol{S}\right | \left |  \boldsymbol{I}_L+
   \left(\boldsymbol{Z}-\boldsymbol{p}\pmb{\hat{\alpha}}^T\right)^{H}\boldsymbol{S}^{-1}
   \left (\boldsymbol{Z}-\boldsymbol{p}\pmb{\hat{\alpha}}^T\right)\right |  .
\ee

Now, with the result (\ref{ZandAlphaHat}),
we can rewrite (\ref{det_R1plusS}) as
\begin{align}\label{detR1plusSfinal}
   \left | \boldsymbol{R}(\pmb{\hat{\alpha}})+\boldsymbol{S}\right |  &\thinspace=\thinspace
  \left |  \boldsymbol{S}\right | \left | \boldsymbol{X}-  \kappa \left ( \boldsymbol{p} \right) \boldsymbol{u}\boldsymbol{u}^{H}\right |  \nonumber \\
   &\thinspace=\thinspace \left |  \boldsymbol{S}\right |
 \left | \boldsymbol{X}\right | \left | \boldsymbol{I}_L- \kappa \left ( \boldsymbol{p} \right) \boldsymbol{X}^{-1}\boldsymbol{u}\boldsymbol{u}^{H}\right |  \nonumber  \\
   &\thinspace=\thinspace \left |  \boldsymbol{S}\right | \left |  \boldsymbol{X}\right |
 \left ( 1- \kappa \left ( \boldsymbol{p} \right)
 \boldsymbol{u}^{H}\boldsymbol{X}^{-1}\boldsymbol{u}\right ) .
   \end{align}
%
From (\ref{det_R0}) and (\ref{detR1plusSfinal}), we get
(\ref{myGLRT}) after a few straightforward steps.

\section{Proof of the CFAR property of the LRT (\ref{ray_based_LRT})}\label{app-c}

We will prove the CFAR property of the LRT (\ref{ray_based_LRT})
%
by showing that the distribution of the maximum eigenvalue $d_{\max}$ of
$\boldsymbol{\Upsilon}$ under the null hypothesis 
does not depend on
the noise covariance matrix $\boldsymbol{C}$.


Let us first decompose
$\boldsymbol{C}$ as 
\begin{equation}\label{eig_decomp_C}
  \boldsymbol{C}=\boldsymbol{U}_C^{H} \pmb{\Lambda}_C \boldsymbol{U}_C ,
\end{equation}
where $\boldsymbol{U}_C$ is an $N \times N$ unitary matrix such that
%
%
$\boldsymbol{U}_C^{-1} = \boldsymbol{U}_C^H$ and $\pmb{\Lambda}_C$ is the
$N \times N$ diagonal matrix composed of the eigenvalues of
$\boldsymbol{C}$.
Next, define 
\begin{equation}
 \boldsymbol{\tilde{Z}} \thinspace=\thinspace \boldsymbol{Q}^{-1}
 \boldsymbol{Z}\label{Z_whitening}
 \end{equation}
 and
 \begin{equation}\label{ZS_whitening}
   \boldsymbol{\tilde{Z}}_S \thinspace=\thinspace \boldsymbol{Q}^{-1} \boldsymbol{Z}_S,
 \end{equation}
where
\begin{equation}\label{C_square_root}
  \boldsymbol{Q} \thinspace=\thinspace \boldsymbol{U}_C^H
  \pmb{\Lambda}_C^{1/2} \boldsymbol{U}_C
\end{equation}
with $\pmb{\Lambda}_C^{1/2}$ denoting the $N \times N$ diagonal matrix
of the square roots of the diagonal elements of $\pmb{\Lambda}_C$: Note that
the diagonal elements of $\pmb{\Lambda}_C$ are positive since $\boldsymbol{C}$
is positive definite. 
Then, under the null hypothesis, we have
$
\boldsymbol{\tilde{Z}} = \left[\boldsymbol{Q}^{-1}\boldsymbol{n}_1,\thinspace
\boldsymbol{Q}^{-1}\boldsymbol{n}_2,
\ldots,\thinspace
\boldsymbol{Q}^{-1}\boldsymbol{n}_L \right]
$
and
$\boldsymbol{\tilde{Z}}_S =
\left[\boldsymbol{Q}^{-1}\boldsymbol{n}_{L+1},
 \right. \linebreak \left.
\boldsymbol{Q}^{-1}\boldsymbol{n}_{L+2},
\ldots, \boldsymbol{Q}^{-1}\boldsymbol{n}_{L+K} \right]$:
%
Thus, for any column of $\boldsymbol{\tilde{Z}}$
and $\boldsymbol{\tilde{Z}}_S$,
we have 
\be
E\left[ \left (\boldsymbol{Q}^{-1}\boldsymbol{n}_j \right )
 \left (\boldsymbol{Q}^{-1}\boldsymbol{n}_j \right)^H
\right] &=& \boldsymbol{Q}^{-1} E\left[\boldsymbol{n}_j
\boldsymbol{n}_j^H\right] \left( \boldsymbol{Q}^{-1} \right )^H \nonumber \\
&=&\boldsymbol{Q}^{-1} \boldsymbol{C} \left( \boldsymbol{Q}^{-1} \right )^H \nonumber \\
&=& \boldsymbol{I}_N ,
\ee
%
%
which implies that the distributions of $\boldsymbol{\tilde{Z}}$
and $\boldsymbol{\tilde{Z}}_S$ do not depend on $\boldsymbol{C}$.

Finally, since $\boldsymbol{\Upsilon}$ can be expressed as 
%
\begin{eqnarray}
  \boldsymbol{\Upsilon}
  &=&
  \boldsymbol{\tilde{Z}}^H \boldsymbol{Q}^H
      \left (\boldsymbol{Q} \boldsymbol{\tilde{Z}}_S
      \boldsymbol{\tilde{Z}}_S^H \boldsymbol{Q}^H
       \right)^{-1}
       \boldsymbol{Q}\boldsymbol{\tilde{Z}}  \nonumber \\
   &=&
      \boldsymbol{\tilde{Z}}^H \left( \boldsymbol{\tilde{Z}}_S
      \boldsymbol{\tilde{Z}}_{S}^H \right)^{-1}
       \boldsymbol{\tilde{Z}}\label{whitening_Z}
\end{eqnarray}
%
from $\boldsymbol{S} = \boldsymbol{Z}_S \boldsymbol{Z}_S^H
=\boldsymbol{Q} \boldsymbol{\tilde{Z}}_S
\boldsymbol{\tilde{Z}}_S^H \boldsymbol{Q}^H$, it is clear that,
under the null hypothesis,
the distribution of the maximum eigenvalue $d_{\max}$ of $\boldsymbol{\Upsilon}$
does not depend on the noise covariance matrix $\boldsymbol{C}$.



\section*{Acknowledgment}
This work was supported by the National Research Foundation of Korea
under Grant NRF-2015R1A2A1A01005868 with funding from the Ministry of Science, Information and
Communications Technology, and Future Planning,
for which the authors wish to express their appreciation.
\ifCLASSOPTIONcaptionsoff
  \newpage
\fi

\end{document}